\documentclass[aps,pre,twocolumn,showpacs,superscriptaddress,noeprint]{revtex4-1}
\usepackage[utf8]{inputenc}
\usepackage{amsmath, amsfonts, amsthm, bbm}
\usepackage{amsthm}
\usepackage{amssymb}
\usepackage{pifont}
\usepackage{graphicx} 
\usepackage{bmpsize}
\usepackage{tikz}
\usepackage{float}

\def\bitcoinA{%
  \leavevmode
  \vtop{\offinterlineskip 
    \setbox0=\hbox{B}%
    \setbox2=\hbox to\wd0{\hfil\hskip-.03em
    \vrule height .3ex width .15ex\hskip .08em
    \vrule height .3ex width .15ex\hfil}
    \vbox{\copy2\box0}\box2}}

\begin{document}

\title{Evolutionary Dynamics of Sustainable Blockchains}

\author{Marco Alberto Javarone}
\email{marcojavarone@gmail.com}
 \affiliation{Centro Ricerche Enrico Fermi, Rome, Italy}
  \affiliation{University College London - Centre for Blockchain Technologies, London, UK}
  
\author{Gabriele Di Antonio}
 \affiliation{Centro Ricerche Enrico Fermi, Rome, Italy} 
  \affiliation{Istituto Superiore di Sanità, Rome, Italy}
 \affiliation{Università degli Studi Roma Tre, Rome, Italy}

\author{Gianni Valerio Vinci}
 \affiliation{Istituto Superiore di Sanità, Rome, Italy}
  \affiliation{Università Roma Tor Vergata, Rome, Italy}

\author{Luciano Pietronero}
 \affiliation{Centro Ricerche Enrico Fermi, Rome, Italy}

\author{Carlo Gola}
\thanks{The usual disclaimer applies}
 \affiliation{Banca d'Italia, Rome, Italy}

\date{\today}
\begin{abstract}
The energy sustainability of blockchains, whose consensus protocol rests on the Proof-of-Work, nourishes a heated debate. The underlying issue lies in a highly energy-consuming process, defined as mining, required to validate crypto-asset transactions.
Mining is the process of solving a cryptographic puzzle, incentivised by the possibility of gaining a reward.
The higher the number of users performing mining, i.e. miners, the higher the overall electricity consumption of a blockchain. For that reason, mining constitutes a negative environmental externality.
Here, we study whether miners' interests can meet the collective need to curb energy consumption. To this end, we introduce the Crypto-Asset Game, namely a model based on the framework of Evolutionary Game Theory devised for studying the dynamics of a population whose agents can play as crypto-asset users or as miners. 
The energy consumption of mining impacts the payoff of both strategies, representing a direct cost for miners and an environmental factor for crypto-asset users. 
The proposed model, studied via numerical simulations, shows that, in some conditions, the agent population can reach a strategy profile that optimises global energy consumption, i.e. composed of a low density of miners. 
To conclude, can a Proof-of-Work-based blockchain become energetically sustainable? Our results suggest that blockchain protocol parameters could have a relevant role in the global energy consumption of this technology. 
\end{abstract}

\maketitle

\section{Introduction}\label{sec:introduction}
The Blockchain~\cite{satoshi01,antonopoulos01} is a distributed ledger able to manage transactions of digital tokens, constituting a form of digital money, without relying on financial institutions serving as trusted third parties. 
The underlying mechanisms of this technology exploit cryptographic protocols, so we refer to digital tokens as crypto-assets. Although all implemented protocols have a role in the functioning of a blockchain, the fundamental one is the Nakamoto consensus protocol.
The latter uses the Proof-of-Work (PoW), i.e. a cryptographic proof that relies on solving a crypto-puzzle for validating transactions and offering solid protection from attacks, such as double-spending~\cite{javarone01}.
More in detail, particular users defined as miners collect crypto-asset transactions not yet recorded in the Blockchain in data structures called blocks. 
The miners' goal is to attach their blocks to the head of the Blockchain. 
To this end, miners have to show a PoW that can be obtained by performing a process called mining.
The first miner to complete the PoW receives a reward, in the form of crypto-asset, which constitutes an economic incentive.
Mining is a competition based on finding a random number (called nonce) whose value depends on the block at the head of the Blockchain and transactions contained in the newly generated block.
Also, mining resembles a lottery, requiring miners to use substantial computational resources to succeed. Then, getting high success probabilities equal high energy consumption in a mining competition, as winning depends on the number of used resources. 
The whole mechanism entails that each new block attached to the chain forms a 'computational shield' preserving transactions recorded in the previously mined blocks.
In summary, miners are fundamental for blockchains based on the PoW. 
On the other hand, the energy consumption of PoW blockchains constitutes a non-trivial issue~\cite{gola01}.
The latter directly impacts the miners' electricity bill and, indirectly, the whole society since a fraction of energy gets wasted for validating transactions.
Even if there are other consensus mechanisms, for instance, the Proof-of-Stake~\cite{saleh01,thin01} and the Proof-of-Authority~\cite{xiao01,zhang01}, the PoW is currently the most used~\cite{aggarwal01}.
Therefore, energy consumption is at the core of a heated debate around the societal impact of blockchains ---see for instance~\cite{keller01,teufel01,vranken01}.
The relevance of the debate motivates this investigation. On one side, we observe miners that compete to get high profits and allow Blockchain to work. On the other, we see a waste of electric energy.
Can the need to curb energy consumption meet the individual interest of miners?
Stimulated by this question, we study the above-described scenario as a social dilemma. To this end, using the framework of Evolutionary Game Theory~\cite{nowak01,moreno01,perc01,perc02,perc03,szolnoki01,szolnoki02,szolnoki03,szolnoki04,javarone02,javarone04} we propose a stochastic evolutionary game named Crypto-Asset Game (CAG).
The proposed game has two strategies, i.e. mining and using crypto-assets. On varying the game conditions, e.g. the miners' reward, we analyse the evolution of the strategy distribution occurring in an agent population playing CAG. In particular, the evolutionary dynamics of CAG entail agents can change their strategy depending on the accumulated payoff via a stochastic process, usually defined strategy revision phase. Hence, a crypto-asset user (or just user hereinafter) can become a miner and vice versa at no cost. 
Rational agents take decisions (i.e. select a strategy) to maximise a profit. We aim to study whether a population of rational agents can reach a strategy profile that minimises global energy consumption.
Finally, let us report that game-based approaches, devised for studying various aspects of blockchains, have been previously proposed, as in~\cite{wu01,li01,liu01,kim01,bai01,teixeira01}.
The remainder of the paper is organised as follows: Section~\ref{sec:model} introduces the Crypto-Asset Game. Section~\ref{sec:results} presents results of numerical simulations. Then, Section~\ref{sec:discussion} highlights the main finding and, eventually, Section~\ref{sec:conclusions} discusses future developments.
\section{The Crypto-Asset Game}\label{sec:model}
The Crypto-Asset Game represents the essential dynamics of a simplified economic system composed of two kinds of agents, i.e. miners and crypto-asset owners that produce and consume crypto-assets, respectively.
More in detail, miners produce crypto-assets through their mining activity, while crypto-asset owners consume crypto-assets by executing transactions~\cite{tessone01}. For the sake of brevity, from now on, we refer to crypto-asset owners as users.
Accordingly, CAG is a two-strategy game whose agents can choose between mining ($m$) and using crypto-assets ($u$). Both strategies are associated with a payoff. The miner payoff includes a reward and an energy cost, while the user payoff includes a gain and a contribution they have to pay. 
The users' gain represents benefits related to using a blockchain, while their contribution represents various costs associated with this technology, as later described. In addition, the overall energy consumption of the blockchain affects the users' payoff as their gain gets multiplied by the miners' density.
Following the above prescriptions, the payoff structure of CAG reads
\begin{equation}\label{eq:payoff}
\begin{cases}
\pi_u = \frac{1}{\rho_m} \frac{N_u}{5} - c_u\\
\pi_m = \langle R \rangle - C
\end{cases}
\end{equation}
\noindent whose parameters have the following meaning. Namely, $\rho_m$ is the density of miners, $N_u$ is the number of users locally involved in a game iteration (later detailed), $c_u$ is the contribution paid by users for participating in the Blockchain system, $R$ is the miners' reward and, eventually, $C$ is the electricity cost of mining.
The miners' reward is taken as $\langle R \rangle = \frac{R}{N_m}$, with $N_m$ total number of miners in the population. This choice is motivated considering that an equal amount of resources in a lottery-like competition (i.e. the mining competition) entails that, on average, a miner gets a reward $R$ divided by the total amount of miners.
Even if the number of miners is related to the amount of energy wasted in a lottery-like competition, PoW-based blockchains cannot work without them. Hence, the presence of miners is mandatory for our economic system, which gets investigated considering a list of assumptions.
Firstly, we assume miners cannot take any profit if there are no crypto-asset transactions (i.e. if there are no users), although real blockchains allow mining also empty blocks. That assumption, combined with the need to have at least one miner in the population, leads to constraining agents' payoff to zero when $\rho_m = 0$ and $\rho_u =0$ (i.e. density of users).
Then, we assume miners have the same computational resources, and users can switch to the mining strategy at no cost.
Lastly, $C$ and $c_u$ are considered constants albeit, in real scenarios, costs and contributions could vary depending on various internal and external factors. 
Let us now briefly clarify the meaning of the contribution $c_u$. To this end, we may refer to the Public Goods Game~\cite{javarone06} (PGG), whose payoff structure is similar to that of CAG.
In the PGG, there are two strategies, namely cooperation and defection. Cooperators make a contribution typically of unitary value, while defectors contribute nothing.
CAG does not include cooperators or defectors, yet the contribution $c_u$ resembles the coin paid by cooperators in the PGG, although it might have a fractional value (see also~\cite{javarone05}).
In this context, $c_u$ represents the adoption costs of new technology and the consequences of using it, e.g. waiting for the validation of transactions.
Now, observing that the ratio between costs and benefits of new technology affects the related diffusion in a population~\cite{hall01}, we highlight the relevance of $c_u$ for studying blockchain dynamics. 
Finally, for the sake of simplicity, we assume the electrical cost $C$ is denominated using the crypto asset of the game, whose market value is constant.
Once clarified the meaning of parameters and values constituting the game payoff, we describe how the agent population evolves.
The evolution process is the same as that usually considered in other evolutionary games, such as the PGG. In particular, a population arranged over a networked structure at the beginning is equally divided between users and miners.
A networked structure entails that agents have a limited number of neighbours and, among them, the number of users corresponds to $N_u$.
After each game iteration, agents undergo a strategy revision phase defined by an updating rule. The latter, in this work, is realised via a Fermi-like distribution~\cite{javarone06} and defined as
\begin{equation}\label{eq:update_rule}
P(s^x \leftarrow s^y) = \frac{1}{1 - e^{-\frac{1}{K} (\pi_y -\pi_x)}}
\end{equation}
\noindent so that the probability the $y$-th agent imposes its strategy, i.e. $s^y$, on the $x$-th agent hinges on the difference in their payoff, namely $\pi_y$ and $\pi_x$, and a parameter $K$ representing the temperature (or noise) of the system.
In real scenarios, some human factors can be represented by the parameter temperature (or noise). 
For instance, the temperature can represent the degree of rationality in individuals, or the uncertainty in economic agents, when they take decisions.
More in detail, the higher the temperature, the higher the probability a strategy revision occurs randomly. Likewise, the lower the temperature, the higher the strategy revision results from a rational choice. 
At high temperatures, i.e. with a lot of noise, the strategy revision phase becomes a coin flip.
The strategy revision phase occurs at each game iteration once agents have played CAG in all groups of belonging. For clarity, considering a population arranged over a bi-dimensional square lattice with continuous boundary conditions, each agent belongs to $5$ different groups identified as follows. 
Then, considering the $x$-th agent, its first group is identified considering that agent with its four direct neighbours. The other four groups are defined considering the $x$-th agent being part of a group whose central agent is one of its neighbours, so the $x$-th agent occupies a side position.
In doing so, agents play the game in five different groups collecting the related payoff, whose total amount is compared with that received by miners via equation~\ref{eq:update_rule}.
\begin{figure*}
    \centering
    \includegraphics[width=7.0in]{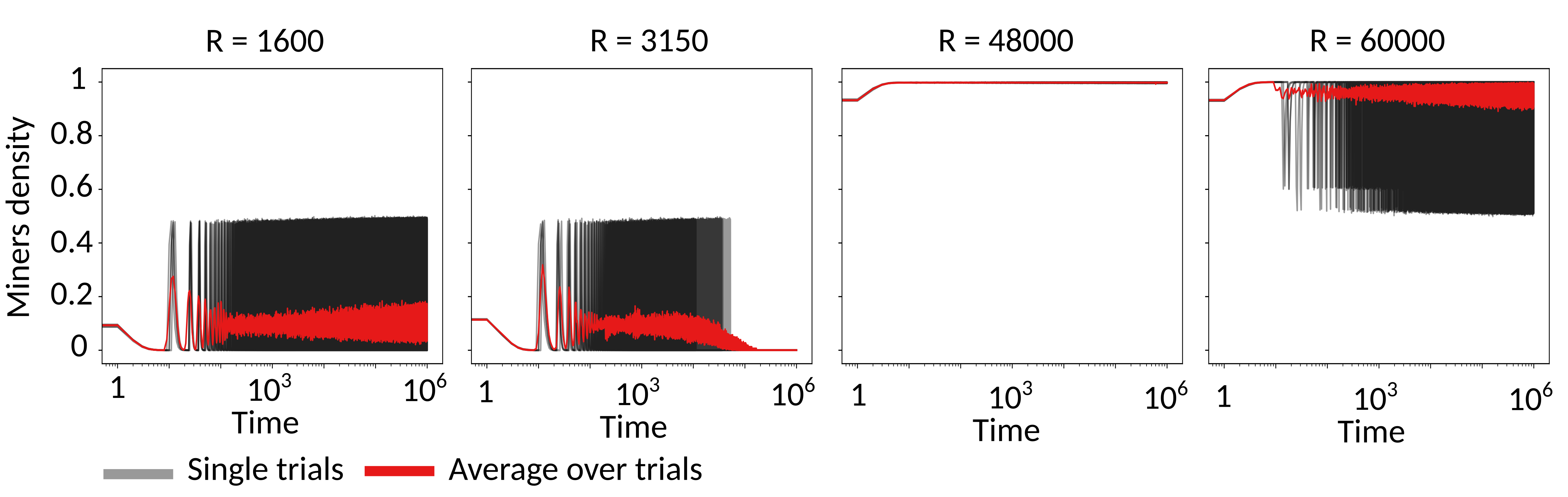}
    \caption{Density of miners over time on varying the miners reward $R$ as indicated on the top of each plot. Each simulation trial lasts $N \times 100$ time steps. Red line indicates the average density value, while a few single trials are reported by black lines. From left to right, results obtained with the following rewards: $R = 1600$, $R = 3150$, $R = 48000$, $R = 60000$}
    \label{fig:figure_1}
\end{figure*}
Finally, there are three possible outcomes, two representing full-order (i.e. full of users and full of miners) and one representing co-existence of strategies.
Assuming a payoff equal to zero when $\rho_m  = 0$ or $\rho_u = 0$ leads to oscillatory dynamics, e.g. in the case of no miners (i.e. $\rho_m  = 0$), the strategy revision phase becomes a coin flip and, on average, half of the population turns the strategy to mining at the next iteration. Hence, the two full-order configurations cannot constitute equilibria of the strategy distribution.
Therefore, our analysis focuses on the emergence of oscillatory behaviours and assessing the phases of strategy co-existence resulting from game dynamics under different conditions.
\section{Results}\label{sec:results}
\begin{figure}
    \centering
    \includegraphics[width=0.96\linewidth]{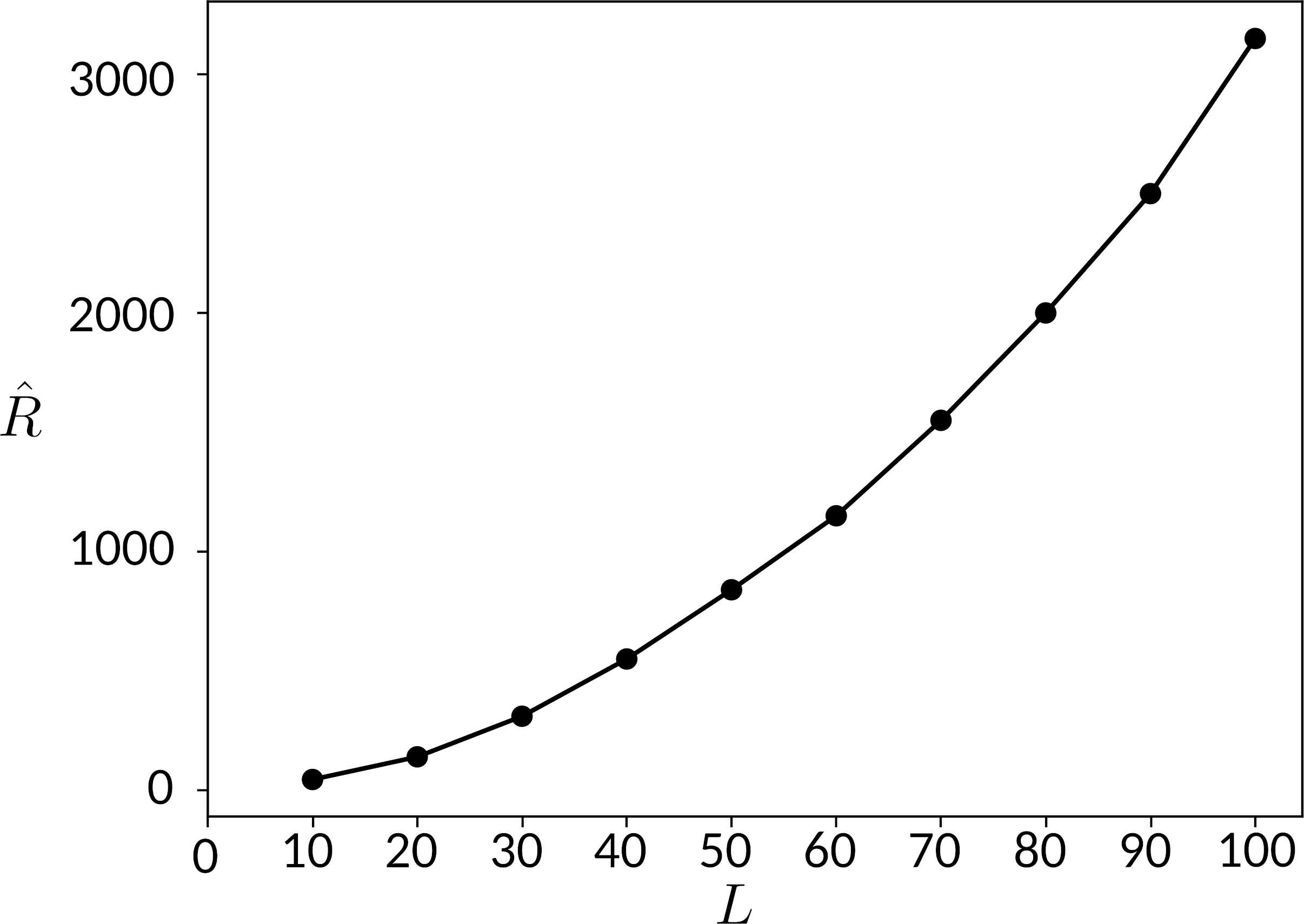}
    \caption{Scaling of the critical reward $\hat{R}$ on varying the population size. The $L$ indicates the size of a (lattice) square side. Thus, for instance, $L = 50$ refers to a population with $L^2 = 2500$ agents.}
    \label{fig:figure_2}
\end{figure}

In this section, we study the Crypto-Asset Game considering populations of different size $N$, from $10^2$ to $10^4$ agents, distributed over a regular square lattice with continuous boundary conditions.
Results are averaged over $100\times N$ attempts, and performed considering a user contribution equal to $c_u = 0.1$ and $K = 0.5$ if not stated otherwise.
The first analysis relates the value of the reward $R$ with the density of miners. 
Figure~\ref{fig:figure_1} presents outcomes for a population of size $100$ obtained at different reward values $R$.
Interestingly, we find that rewards smaller than $R = 3100$ in a population with $10^4$ agents, lead to steady-states characterised by an oscillatory behaviour. More in detail, as shown in Figure~\ref{fig:figure_1}, a low reward entails that almost half of the population changes continuously strategy over time, albeit the average behaviour suggests a configuration more stable with only few miners. 
Then, increasing the reward $R$ up to $3150$ for the same population, after an initial phase characterised by several oscillations like in the previous case, the agents reach a steady state with a very low density of miners. 
Higher values of $R$ lead to higher densities of miners till reaching a unitary miners density for $R = 48000$. As soon as $R > 50000$ an oscillatory behaviour, conceptually similar to the one observed at very low $R$ values, emerges. In the last case, almost the whole population oscillates between a miners density equal to $1$ and about $0.6$.
Such analysis is then repeated on varying the size of the population and allows us to identify threshold values in range of the reward $R$ for each specific case.
In figure~\ref{fig:figure_2} we show the threshold values of reward, we name critical rewards, on varying the population size.

\begin{figure}
    \centering
    \includegraphics[width=0.96\linewidth]{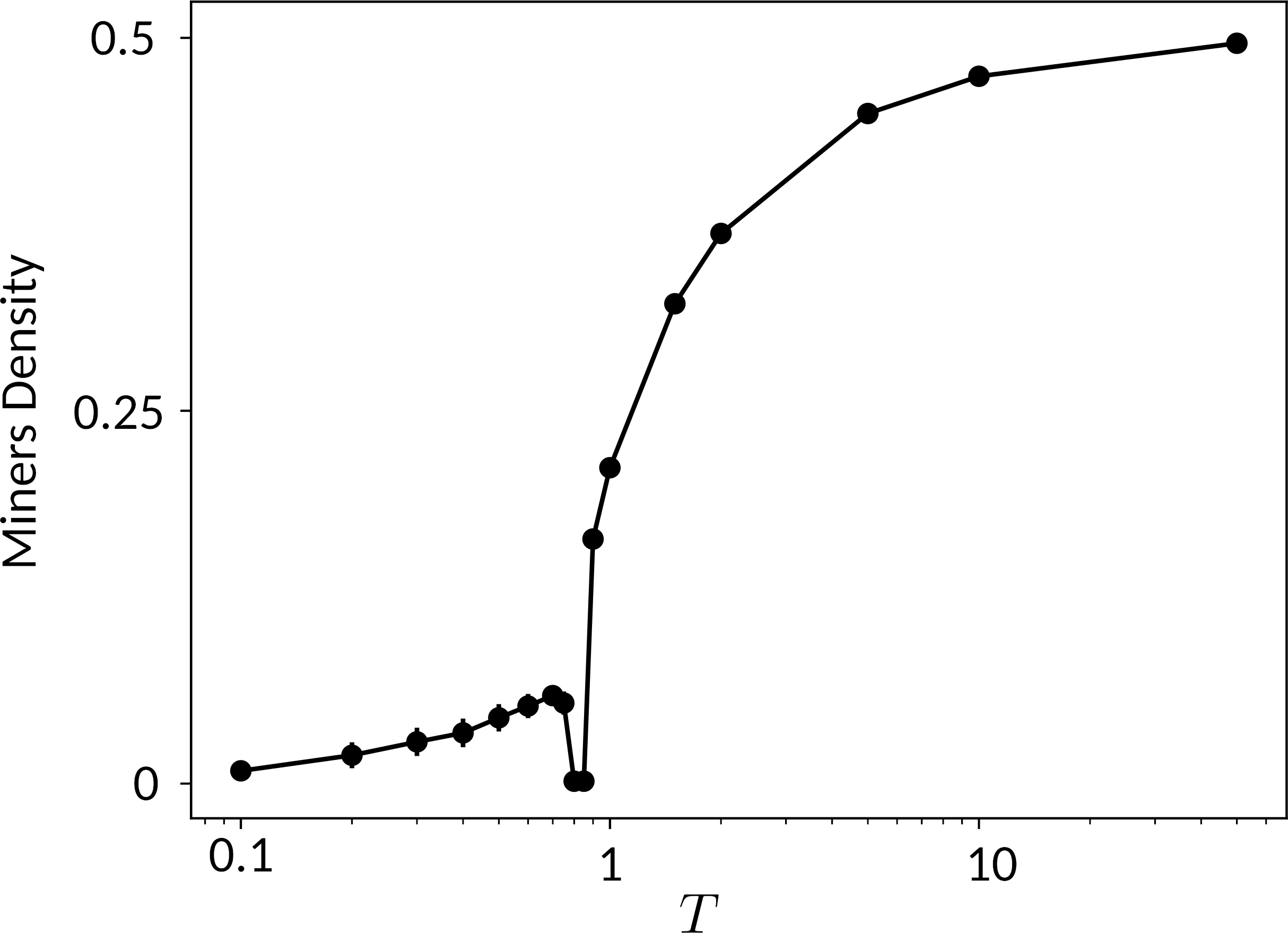}
    \caption{Density of miners at the steady-state (i.e. after $100\times N$) on varying the temperature $T$ in a population composed of $N = 2500$ agents, playing with a reward $R= 900$. The standard error of the mean is very small on all samples, remaining hidden inside the circles on the curve.}
    \label{fig:figure_3}
\end{figure}

\begin{figure*}
    \centering
    \includegraphics[width=\linewidth]{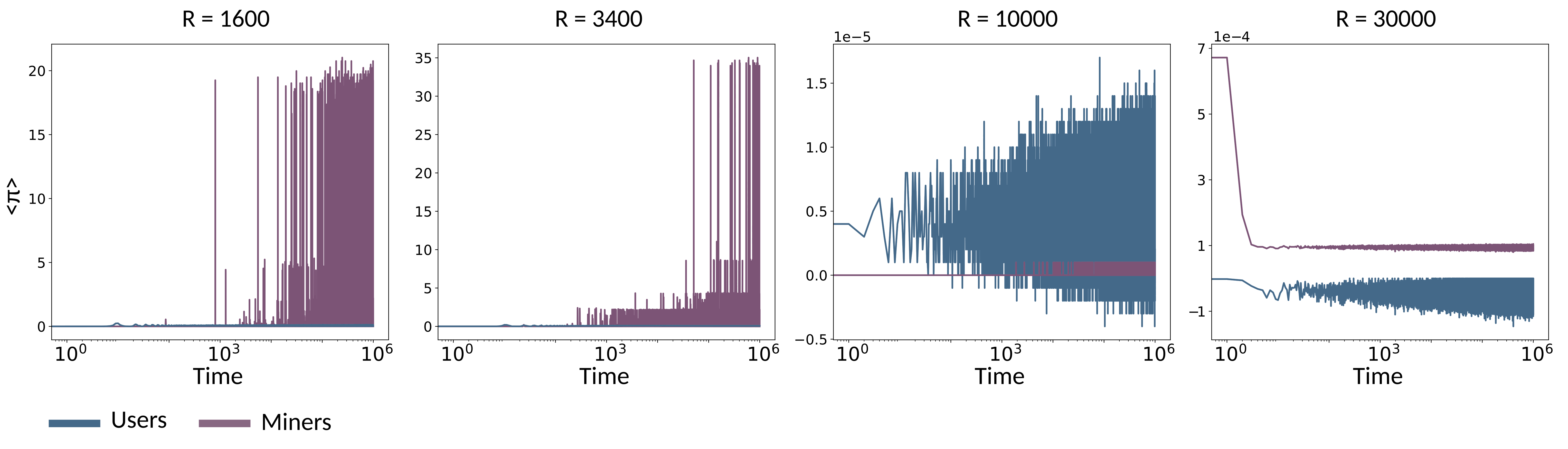}
    \caption{Average payoff  $\langle \pi \rangle$ over time, obtained by users (blue) and miners (purple line) on varying the reward $R$ in a population with $N = 10^4$ agents.}
   \label{fig:figure_4}
\end{figure*}

\begin{figure}
    \centering
    \includegraphics[width=0.96\linewidth]{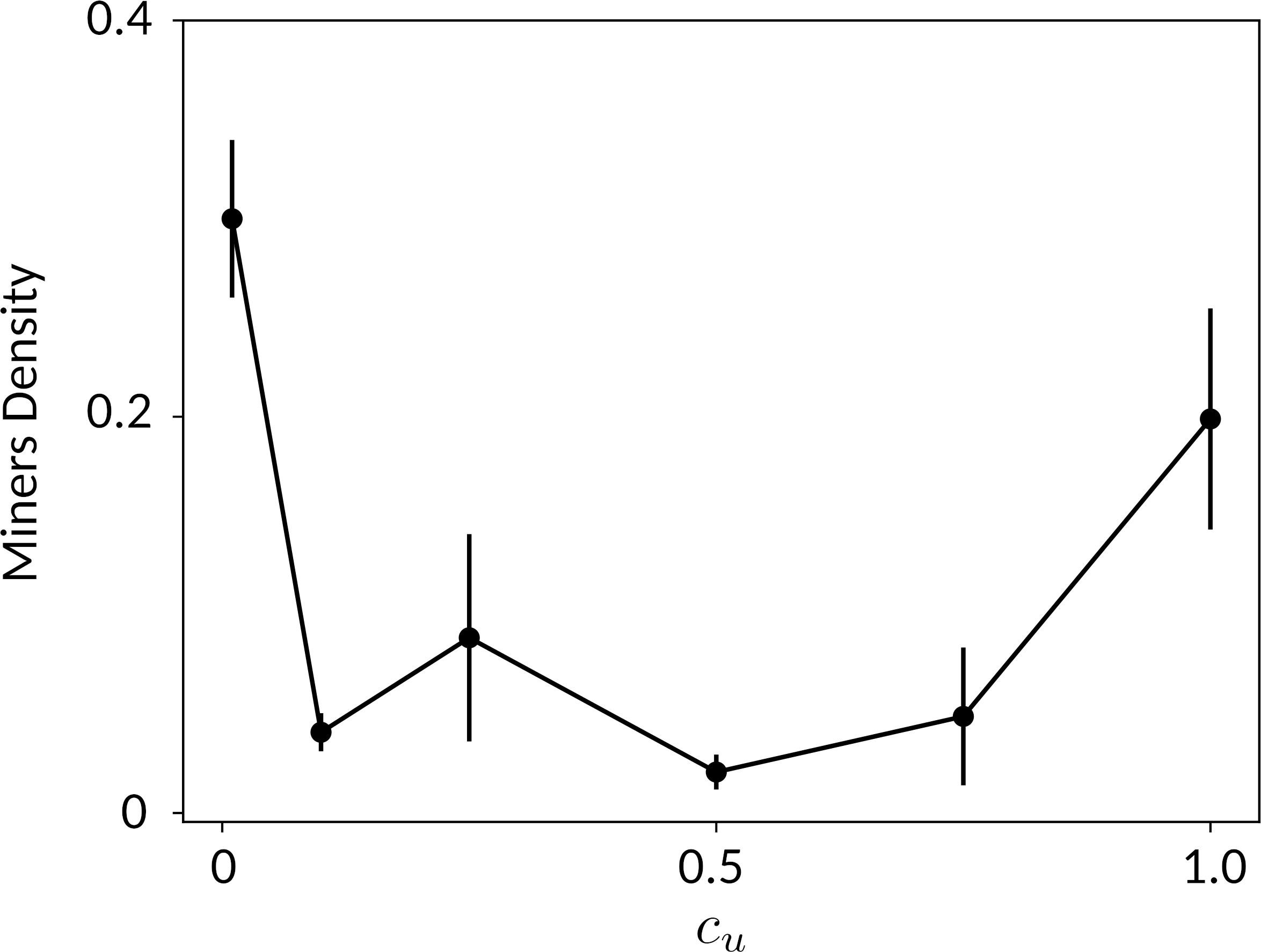}
    \caption{Miners density on varying the user contribution $c_u$ in a population with $N= 2500$ agents. Error bars report the standard error of the mean for each sample.}
   \label{fig:figure_5}
\end{figure}

According to the model, CAG has several parameters that can be investigated to assess their effect on the agent population, with particular interest for the density of miners.
Starting with the temperature $K$, we consider a population with $2500$ agents, to analyse the outcomes at different degrees of rationality.
Such analysis is performed with special attention to two ranges, the first one for $K \in [0,1]$ and the second range for $K \in [1,50]$.
Previous investigations on dilemma games, as the Prisoners' dilemma and the Public Goods Game, reported that for the range of low temperatures the density of cooperators increases as the temperature increases from $0$ to $0.5$ and then the beneficial effect of the temperature on supporting cooperation reduces~\cite{javarone03}. Then, as reported in~\cite{javarone06}, increasing the temperature the dynamics of an evolutionary game resembles that of the voter model~\cite{liggett01}. Namely, too high temperatures entail agents change strategies by a process equivalent to a coin flip.
In this case, we cannot map cooperation to the use of a token and defection to mining, because as described before, proof-of-work-based blockchains need miners, while populations are perfectly healthy without defectors.
The result of this analysis is shown in figure~\ref{fig:figure_3}.
We find that for temperatures between $0$ and $1$ the behaviour is similar to the one we observe in social dilemma games. Notably, increasing the temperature in that range, the density of miners reaches a small peak with around $T = 0.75$. 
Then, as the temperature goes towards high values, the density of miners increases till $0.5$, i.e. as expected, as soon as agents reduce their rationality, the game resembles a voter model, and both strategies survive in the population almost with the same density (i.e. $0.5$). 
The agents' payoff is a relevant component of a game. Then, as shown in Figure~\ref{fig:figure_4}, we study the average payoff of both strategies by varying the reward in a population with $N = 10^4$ agents. Results indicate that miners can reach the highest payoff, on average, as the reward is slightly higher than the critical value for that population size ---see the second plot from the left-hand side in Figure~\ref{fig:figure_4}. Then, increasing the reward, the emerging configurations lead both strategies to become poorly convenient. 
That justifies the behaviour we observed analysing the density of miners on varying the reward $R$. When only very few miners survive, miners and users receive the highest possible payoff they can reach, and the population benefits from a virtuous (in terms of energy consumption) steady state.
We conclude the section studying how the user contribution $c_u$ affects the miners density ---see Figure~\ref{fig:figure_5}.
As reported, a too low user contribution entails the density of miners is not optimal as too many miners survive in the population. That result is neither trivial nor expected, and it resembles the behaviour we observe for too small rewards $R$, i.e. smaller than a critical value $\hat{R}$ (for each population size). 
Finally, as the $c_u$ gets close to $1$, the density of miners slightly increases. In summary, a user contribution ranging from $0.1$ to $0.75$ supports optimised strategy distributions. 
\section{Discussion}\label{sec:discussion}
Energy sustainability is a critical issue of Proof-of-Work-based blockchains and feeds a heated debate. 
This class of blockchains requires computational resources for verifying transactions, which might lead to high energy consumption.
In an economic system composed of rational agents, the energy sustainability of a PoW blockchain can be treated as a social dilemma. 
In the proposed model, miners and crypto-asset users interact by playing a game and getting a payoff, depending on their strategy.
The miners' payoff includes a reward whose market value~\cite{kristoufek01,blau01,javarone07} can make mining highly profitable.
Thus, the resulting dilemma on one end of the scale has the individual interest of miners and, on the other, the collective interest in curbing electricity consumption.
The global energy consumption of a blockchain is linked to the nature of the mining competition. Namely, the higher the number of resources used for mining, the higher the average reward miners get over time.
Increasing the computational power entails paying higher electricity costs. Yet, as in lotteries, whose chances of success increase by buying more tickets, chances of success in the mining competition increase by using more computational resources. 
Sadly, albeit miners pay their electricity bills, such a competition entails society wastes a fraction of energy.
The overall scenario, characterised by the above-described dilemma, can be studied by the framework of Evolutionary Game Theory. 
Therefore, motivated by all these observations, we propose a two-strategy game called the Crypto-Asset Game. 
The complexity of an economic system leads us to make several assumptions in the proposed model. For instance, miners and users can change strategy at no cost, all miners have the same computational resources, the price of electricity is constant, and the value of the crypto-asset in the market does not change.
In summary, CAG simplifies real scenarios. Notwithstanding, we deem CAG captures the essential dilemmatic aspects affecting rational individuals using blockchains.
The results of our model, obtained via numerical simulations, show a rich spectrum of possible steady states that the agent population can reach. More specifically, depending on the game parameters, the agent population can converge toward steady-states characterised by a small number of miners or toward oscillatory behaviours characterised by a fraction of agents continuously switching between the two strategies. Such an oscillatory behaviour agrees with previous studies related to energy aspects of PoW blockchains~\cite{fiat01}.
We recall that the parameters of CAG represent the protocol parameters of PoW blockchains.
Finding steady-state configurations corresponding to optimal strategy distributions, i.e. with a limited number of miners in the population, suggests that the protocol parameters may have a relevant role in blockchain sustainability.  
Among such parameters, the miners' reward is probably the most important.
More specifically, results indicate that rewards too low and too high are detrimental to blockchain energy sustainability, as the density of miners is relatively high in both cases.
Remarkably, we identify a range of values that rewards can take to ensure the evolution of a population towards a virtuous strategy distribution. Such a range of values depends on the population size. 
The smaller the population size, the smaller the value a reward can take for being beneficial (for the society itself). 
In real blockchains, the miners' reward reduces over time~\cite{antonopoulos01}, so the mining activity gets preserved by other mechanisms, such as the transaction fees that users can add (voluntarily) to their transactions.
To conclude, we observe that the definition of a range of beneficial rewards appears, in terms of sustainability, in agreement with real-world blockchain dynamics. Notably, as more users adopt and use a crypto-asset, the miners' reward gets adapted according to various factors, such as those mentioned before. 
That does not entail rewards change to comply with environmental needs however, the miners' reward and the community size in a real blockchain do not seem uncorrelated parameters.
\section{Conclusions}\label{sec:conclusions}
In summary, we deem that the proposed model captures fundamental aspects of PoW blockchains. 
Notwithstanding, several model features constitute relevant elements for future research. 
Just to cite a few, elements deserving further attention relate to the introduction of switching costs for representing the efforts required for mining~\cite{thum01}, the volatility of the crypto-asset value in the market~\cite{aalborg01}, and other mechanisms, such as the halving of the miners' reward. In addition, in light of the structure of transaction networks~\cite{kondor01,tessone01,tessone02,tessone03} we observe in various cryptocurrencies, the interaction topology of the proposed model deserves additional investigations. 
For instance, to study whether network structures~\cite{estrada01}, such as scale-free networks and small-world networks, can affect the strategy profile of a population playing the Crypto-Asset Game, as observed in social dilemmas~\cite{santos01,santos02,cardillo01,amaral01}. 
Let us conclude by remarking that the optimal steady states reached by agents in some game conditions suggest that protocol designers may have a relevant role in the energetic impact of PoW-based blockchains. 
In particular, the miners' rewards could get designed to promote virtuous user behaviours, making, in the long run, this technology sustainable and environmentally-friendly.

\end{document}